\documentclass[12pt]{article}
\usepackage{geometry}
\usepackage{amsmath}
\newcommand{\unit}[1]{\ensuremath{\, \mathrm{#1}}}

\begin{document}

\title{Searching for spacetime granularity: analyzing a concrete experimental setup}

\author{Yuri Bonder\footnote{yuri.bonder@nucleares.unam.mx} and Daniel Sudarsky\footnote{sudarsky@nucleares.unam.mx}\\Instituto de Ciencias Nucleares\\
Universidad Nacional Aut\'{o}noma de M\'{e}xico
\\ A. Postal 70-543, Distrito Federal 04510, M\'exico}

\maketitle

\section*{Abstract}
In this work we show that the spin pendulum techniques developed by the E\"ot-Wash group could be used to put very stringent bounds on the free parameters of a Lorentz invariant phenomenological model of quantum gravity. The model is briefly described as well as the experimental setup that we have in mind.

\section{Introduction}

It is generally believed that General Relativity may be a ``low energy'' description of spacetime. In particular, close to the Planck scale a non-trivial spacetime structure may replace the smooth pseudo-Riemannian one. This regime was for a long time deemed unreachable, but lately there has been a reconsideration of that pessimistic outlook leading to searches for traces of that quantum gravity realm. By far the most popular way for looking for that kind of effects is through violations of Lorentz symmetry \cite{Initial-LIV}. However, there are very stringent experimental bounds limiting that possibility \cite{Bounds}. In addition, the authors of Ref. \cite{Collins} showed that, if a discrete structure of spacetime is associated with a preferential reference frame, its effects would be generically ``magnified'' by radiative corrections to a point where they would have been already observed. 

These arguments lead us to consider that if spacetime has any non-trivial structure at a microscopic scale, it must respect Lorentz symmetry. It is of course very hard to picture how can something be granular and, at the same time, be Lorentz invariant. However, we believe that it is possible study phenomenologically the effects that a Lorentz respectful granularity of spacetime have over matter fields without specifying the way spacetime behaves microscopically. A model with these characteristic was first proposed in Ref. \cite{QGP} and in subsequent works \cite{QGP2,unambiguous} this model was refined rendering it free of the ambiguities that afflicted the original proposal. This model is briefly described in the next section. The rest of the paper is structured as follows: First we describe the experimental setup for an experimental proposal to search for the effects. Then we present the calculation of the effective Hamiltonian appropriate for this particular experimental scenario and we finish with some concluding remarks.

\section{The essence of the model}

The phenomenological model assumes that, if spacetime has a granular micro-structure, it must respect Lorentz invariance. We argue, by following an analogy, that matter fields can be sensible to this granularity through an interaction with spacetime curvature. The analogy is the following: Imagine that we have the task to fill the floor of a room with square blocks (mosaics). If the room has a symmetry compatible with the symmetry of the blocks, \textit{i.e.}, straight walls making angles of $\pi/2$, this can be achieved by simply putting one block next to the other. However, if a wall is curved, we will have to overlap some of the blocks to fill the area close to this wall. We believe that something analogous may happen in spacetime. Consider a spacetime that is built with blocks. Based on the arguments given in the last section we believe that these blocks respect Lorentz symmetry. In regions of spacetime having this symmetry, namely, flat spacetime regions, all the blocks fit without overlapping. On the other hand and according to the analogy, in a curved spacetime region its building blocks ``overlap'' and this is what matter fields may detect. 

Motivated by the simple analogy, we take the view that the granular structure of spacetime may be revealed by an interaction of matter fields and spacetime curvature. As the matter self-interactions are not relevant at the phenomenological level, we are only concerned with the part of the curvature connected with sources away from the probing particles. That part of the curvature is the Weyl tensor. Thus, we have to find a minimally suppressed coupling term of Weyl and the probing particles, which we take to be electrons.

As is discussed in Ref. \cite{QGP}, the most obvious minimally suppressed terms vanish, therefore, a less natural looking coupling term must be constructed. The proposal is to use the eigen-forms and eigen-values of two self-adjoint operators built from the Weyl tensor and couple them in a non-trivial manner with $\bar{\psi}\gamma^\mu \gamma^\nu \psi$, where $\psi$ represent the electron field and $\gamma^\mu$ are the Dirac matrices. 

Essentially the idea is to write the Weyl tensor as a $6 \times 6$ matrix by numbering its components with anti-symmetric pairs of spacetime indexes. This matrix, when using Riemann normal coordinates about the point of interest, has the generic form 
\begin{eqnarray} \label{Weylgral}
\left(\begin{array}{cccc}
\mathbf{ A}&\mathbf{ B}\\
-\mathbf{B}&\mathbf{ A}
\end{array} \right),
\end{eqnarray}
where $\mathbf{A}$ and $\mathbf{B}$ are $3 \times 3$ real traceless symmetric matrices \cite{Weyl matrix form}. Then the coupling term can be written in terms of the (real) eigen-values and the ($3$ dimensional) eigen-vectors of $\mathbf{A}$ and $\mathbf{B}$. Let $\vec{a}^{(l)}$, $\vec{b}^{(l)}$ be two orthonormal triads of vectors\footnote{Latin indexes run from $1$ to $3$.} such that
\begin{eqnarray}\label{eq de eigenvalores A y B}
{A_i}^ja^{(l)}_j=\alpha^{(l)} a^{(l)}_i,\qquad
{B_i}^jb^{(l)}_j=\beta^{(l)} b^{(l)}_i,
\end{eqnarray}
The effective Hamiltonian of the model (obtained by using the result of Ref. \cite{NRH} which is valid in the framework of the Standard Model Extension), in the reference frame where the probing polarized electrons are at rest, takes the form\footnote{In this work we use an improved model in comparison with Ref. \cite{unambiguous} where the $sign$ function is replaced by its argument. This is because $sign$ is discontinuous and leads to non-physical effects. Of course, with this modification the scheme is still well defined and unambiguous.}
\begin{eqnarray}\label{HNRs}
\mathcal{H}_{eff}=&\sum_{l,m=1}^3 \Delta\xi^{(l,m)} |\alpha^{(l)}|^{1/4}|\beta^{(m)}|^{1/4} \left(\frac{|\alpha^{(l)}|^{1/2}}{M_{Pl}}\right)^{c^{(+,l)}}\left(\frac{|\beta^{(m)}|^{1/2}}{M_{Pl}}\right)^{c^{(-,m)}} 
\nonumber\\ &
\vec{a}^{(l)}\cdot \vec{b}^{(m)} \vec{\sigma}\cdot\vec{a}^{(l)}\times \vec{b}^{(m)},
\end{eqnarray}
where $\Delta\xi^{(l,m)}$ and $c^{(\pm,l)}$ are the free parameters of the model\footnote{The parameters $\Delta\xi^{(l,m)}$ correspond with $\xi^{(+,-,l,m)}-\xi^{(-,+,m,l)}$ of Ref. \cite{unambiguous}}, $\sigma_i$ are the Pauli matrices, $M_{Pl}$ is Planck mass and $sign$ is a function that is $1$, $0$ or $-1$ if its argument is positive, zero or negative, respectively. The arrow represents $3$-dimensional vectors while $\cdot$ and $\times$ are the usual (Euclidean) scalar and vector products. Observe that in the units where $c=\hbar=1$, which are the units that are being used, $\alpha^{(l)}$ and $\beta^{(l)}$ have mass dimension $2$ and thus $\mathcal{H}_{eff}$ has dimensions of mass.

\section{Experimental scenario}

The presence of the Pauli matrices in the effective Hamiltonian (\ref{HNRs}) indicates that non-polarized matter is not sensible to the effect described above. This makes its search difficult since most of the bulk matter found in Nature is not polarized. Moreover, polarized matter has normally a magnetic moment and the resulting magnetic effects would dwarf the effects we are interested on.

In Ref. \cite{unambiguous} we were able to put some bounds on the free parameters of the model by using data from a Hughes-Drever-like experiment \cite{datos} where certain spectral lines are monitored with a one-year period. According to the model described in the last section this spectral lines would depend on the gravitational field where the probing particles are. Since the Sun's gravitational effect on Earth has a one-year period (due to the Earth's ellipticity), according to the model, a modulation of the spectral lines with this period is expected. The absence of this variation up to the experimental accuracy allowed us to put bounds on the free parameters of the model.

The purpose of this paper is to show that it is possible to do a different type of experiment to test the model at hand. In addition, that this experiment can improve significantly the bounds on the free parameters, and optimistically, to find a signature the conjectured effect. The idea is to use the ``spin pendulum'' developed by the E\"ot-Wash group \cite{Adelberger} which has an effective polarization of roughly $10^{23}$ electrons but a negligible magnetic moment. The spin pendulum can be placed on a torsion balance where a torque exerted on it can be measured with an amazing precision. The concrete idea is to produce a gravitational environment with controlled sources at the Laboratory. In particular, two dense masses can be placed symmetrically with respect to the spin pendulum on an horizontal plane at the Lab, successively at different orientations. As the effect involves also the frame dragging associated with the Earth's angular momentum, the different orientations would result in a differential response of the balance. The calculation of the effect in that situation is presented on the next section.

\section{Calculations for the experimental setup}

In this section we present the calculations for the model in the experimental setup described at the end of the last section. The analysis focuses at first on a single electron. We use normal Riemannian coordinates associated with the location of the electron and the instant where the measurement is performed. Moreover, the spatial part of the coordinates is taken to be right-handed having the $z$ coordinate pointing up and the $y$ pointing south. Note that at the origin, where all the objects that we calculate are evaluated, the metric has the usual Minkowski form\footnote{We use the $+2$ signature for the metric.}. (The model is fully covariant thus we use coordinates that simplify the analysis). 

Provided that gravity is ``weak'', the linearized approximation for the gravitational field is used. In this regime (and using the standard gauge, see \cite{Wald}), the dominant part of $\mathbf{A}$ and $\mathbf{B}$ take the form
\begin{equation}
\label{AB}{A_i}^j = \partial_{i} \partial^{j}\Phi_N,\qquad
{B_i}^j=\left[\partial_i curl(\vec{\Pi})^j +\partial^j curl(\vec{\Pi})_i\right],
\end{equation}
where 
\begin{equation}
\Phi_N(\vec{x},t)= G \int \frac{\rho(\vec{x}', t)}{|\vec{x}-\vec{x}'|}d^3x' , \qquad
\vec{\Pi}(\vec{x},t)= G \int \frac{\vec{p}(\vec{x}', t)}{|\vec{x}-\vec{x}'|}d^3x',
\end{equation}
$G$, $\rho$ and $\vec{p}$ representing Newton's constant, and the matter and momentum density of the gravitational sources, respectively. Note that $\Phi_N$ is the usual Newtonian gravitational potential.

The matrix $\mathbf{A}$ corresponding to the sources at the Laboratory (denoted with a subindex $L$) is calculated using the first equation (\ref{AB}). We assume that the sources are point-like mass $m$ and are located at $(d \cos \theta,d \sin\theta, z_0)$ and $(-d \cos \theta,-d \sin\theta, z_0)$, where $0\leq \theta< \pi$. We then find,
\begin{small}\begin{eqnarray}
\mathbf{A}_L= \frac{G m d^2}{(d^2+z_0^2)^{5/2}} \left(
\begin{array}{ccc}
1+3 \cos (2 \theta )-2z_0^2/d^2 & 3 \sin (2 \theta ) & 0 \\
3 \sin (2 \theta ) & 1-3 \cos (2 \theta )-2 z_0^2/d^2& 0 \\
 0 & 0 & -2+4z_0^2/d^2
\end{array}\right).
\end{eqnarray}\end{small}
It can be seen that the matrix $\mathbf{B}$ depends on the ``movement'' of the gravitational sources, therefore, in this reference reference frame, $\mathbf{B}_L=0$.

Note that in order to have a non-trivial effective Hamiltonian we need both, $\mathbf{A}$ and $\mathbf{B}$, to be non-zero, thus, when calculating the effect of the Earth gravitational field we also need to take into the account its rotation. For this purpose instead of using equations (\ref{AB}) we follow a different strategy. We start with the Kerr metric and we calculate the linearized Weyl tensor. Then we translate its components to the reference frame that is instantaneously at rest with respect to the probing particle. In this way we obtain:
\begin{eqnarray}
\mathbf{A}_\oplus=\frac{GM}{R^3}\left(
\begin{array}{ccc}
 -1 & 0 & 0 \\
 0 & -1 & 0 \\
 0 & 0 & 2
\end{array}
\right), \quad
\mathbf{B}_\oplus=\frac{3 J G}{R^4}\left(
\begin{array}{ccc}
 \cos \theta_L & 0 & 0 \\
 0 & \cos \theta_L & - \sin \theta_L \\
 0 & - \sin \theta_L & -2 \cos \theta_L
\end{array}
\right),
\end{eqnarray} 
where $\theta_L$ represents the co-latitude of the Laboratory, the subindex $\oplus$ is used to denote the Earth, and $R$, $M$ and $J$ stand respectively for the Earth's radius, mass and angular momentum.

Since we are working in the linearized regime, we have
\begin{eqnarray}
\mathbf{A}=\mathbf{A}_\oplus+\mathbf{A}_L, \qquad
\mathbf{B}=\mathbf{B}_\oplus.
\end{eqnarray}
The eigen-values are (see equation (\ref{eq de eigenvalores A y B}) to recall the notation):
\begin{eqnarray}
\alpha^{(1)}= \frac{2 G M}{R^3}-\frac{2 G m \left(d^2-2 z_0^2\right)}{\left(d^2+z_0^2\right)^{5/2}}, & \beta^{(1)}= \frac{3 G J}{ R^4} \cos\theta_L,\\
\alpha^{(2)}=-\frac{GM}{R^3}-\frac{2 Gm}{\left(d^2+z_0^2\right)^{3/2}},& \beta^{(2)}= \frac{3 G J}{2 R^4}\left(-\cos\theta_L -\sqrt{5\cos^2\theta_L+4}\right), \\
\alpha^{(3)}= -\frac{GM}{R^3}+\frac{2 G m(2 d^2 -z_0^2)}{\left(d^2+z_0^2\right)^{5/2}},&
\beta^{(3)}= \frac{3 G J}{2 R^4}\left(-\cos\theta_L +\sqrt{5\cos^2\theta_L+4}\right).
\end{eqnarray}
With corresponding orthonormal eigen-vectors
\begin{eqnarray}
\vec{a}^{(1)}= (0,0,1),&
\vec{b}^{(1)}= (1,0,0),\\
\vec{a}^{(2)}=|\cos\theta| (-\tan \theta ,1,0),&
\vec{b}^{(2)}=(0,-3 \cos \theta_L+S,2 \sin\theta_L)/N_+,\\
\vec{a}^{(3)}=|\sin\theta|(\cot\theta ,1,0),&
\vec{b}^{(3)}=(0,-3 \cos \theta_L-S,2 \sin\theta_L)/N_- , 
\end{eqnarray}
where
\begin{equation}
S=\sqrt{5 \cos^2\theta_L+4},\qquad N_{\pm}=\sqrt{10 \cos^2\theta_L \mp 6S\cos\theta_L+8}.
\end{equation}
Observe that in this derivation we assume $\sin \theta_L>0$, thus, the expressions for $\vec{b}^{(2)}$ and $\vec{b}^{(3)}$ are valid in any place on Earth excepting its poles.

We turn back to the effective Hamiltonian (\ref{HNRs}). Assume that the effective polarization of the spin pendulum is equivalent to the polarization of $N(\approx 10^{23})$ electrons pointing in the direction $\hat{n}=(\cos\varphi,\sin\varphi,0)$. Thus, the expectation value of the total Hamiltonian is
\begin{eqnarray}\label{E}
E_e=\left<\mathcal{H}_{eff}\right>=&\sum_{l,m=1}^3 N \Delta\xi^{(l,m)} |\alpha^{(l)}|^{1/4}|\beta^{(m)}|^{1/4} \left(\frac{|\alpha^{(l)}|^{1/2}}{M_{Pl}}\right)^{c^{(+,l)}}\nonumber\\ &\left(\frac{|\beta^{(m)}|^{1/2}}{M_{Pl}}\right)^{c^{(-,m)}} \vec{a}^{(l)}\cdot \vec{b}^{(m)} (\hat{n}\cdot\vec{a}^{(l)}\times \vec{b}^{(m)}),
\end{eqnarray}
where for each electron state we have $\left< \vec{\sigma} \right>= \hat{n}$. The total energy and torque exerted on the torsion balance can be written as
\begin{eqnarray}
E=-N \hat{n} \cdot \vec{w},\qquad
\vec{T}=N \hat{n} \times \vec{w}, 
\end{eqnarray}
for a certain vector $\vec{w}$. When comparing this expression for the total energy $E$ with equation (\ref{E}) we can read off $\vec{w}$, which can then be inserted in the expression for the torque. The quantity to be measured with the spin pendulum is the $z$ component of the torque, which according to the model is given by
\begin{eqnarray}\label{Tz}
T_z=& \sum_{l,m=1}^3 \Delta\xi^{(l,m)} N |\alpha^{(l)}|^{1/4}|\beta^{(m)}|^{1/4} \left(\frac{|\alpha^{(l)}|^{1/2}}{M_{Pl}}\right)^{c^{(+,l)}}\left(\frac{|\beta^{(m)}|^{1/2}}{M_{Pl}}\right)^{c^{(-,m)}}\nonumber\\& \sin\theta_L(f^{(l,m)}_y \cos\varphi -f^{(l,m)}_x \sin\varphi)/S, 
\end{eqnarray}
where $f^{(l,m)}_i=S \vec{a}^{(l)}\cdot \vec{b}^{(m)} (\vec{a}^{(l)}\times \vec{b}^{(m)})_i/\sin \theta_L$. The values of $f^{(l,m)}_x$ and $f^{(l,m)}_y$ are presented in the following table:
\begin{center}
 \begin{tabular}{| c | c || c | c |}
 \hline
 $l$ & $m$ & $f^{(l,m)}_x$ & $f^{(l,m)}_y$\\ 
 \hline 
 $1$ & $1$ & $0$ & $0$ \\ 
 $1$ & $2$ & $-1$ & $0$ \\
 $1$ & $3$ & $1$ & $0$ \\ 
 $2$ & $1$ & $0$ & $0$ \\
 $2$ & $2$ & $\cos^2\theta$ & $\cos\theta \sin\theta$ \\ 
 $2$ & $3$ & $-\cos^2\theta$ &  $-\cos\theta \sin\theta $\\
 $3$ & $1$ & $0$ & $0$ \\ 
 $3$ & $2$ & $\sin^2\theta$ & $-\cos\theta \sin\theta $ \\
 $3$ & $3$ & $-\sin^2\theta$ &  $\cos\theta \sin\theta $\\ 
 \hline
 \end{tabular}
\end{center}

If we denote by $\Delta T_z$ the experimental precision for measuring $T_z$, then, if the conjectured signal is not observed, the result of the experiment would be that the absolute value of the right-hand side of equation (\ref{Tz}) is smaller than $\Delta T_z$. If in addition we assume, as is customary, that there are no fortuitous cancellations among the terms in this sum we can conclude
\begin{equation}\label{cotas}
|\Delta\xi^{(l,m)}| |\alpha^{(l)}|^{1/4}|\beta^{(m)}|^{1/4} \left(\frac{|\alpha^{(l)}|^{1/2}}{M_{Pl}}\right)^{c^{(+,l)}}\left(\frac{|\beta^{(m)}|^{1/2}}{M_{Pl}}\right)^{c^{(-,m)}}<\frac{\Delta T_z}{N},
\end{equation}
where we use that $\sin\theta_L/S \leq 1/2$, $|f^{(l,m)}_x|\leq 1$ and $|f^{(l,m)}_y|\leq 1$. Note that relation (\ref{cotas}) is only meaningful for terms with non-vanishing $f^{(l,m)}_x$ and $f^{(l,m)}_y$. 
\section{Conclusions}

In this work it is shown that the spin pendulum developed by the E\"ot-Wash group can be used to test the Lorentz invariant phenomenological model of quantum gravity presented in \cite{QGP,QGP2,unambiguous}. The E\"ot-Wash group can measure the torque on the spin pendulum with a precision of $10^2 \unit{eV}$ \cite{Adelberger2} and since the spin pendulum has an effective polarization equivalent to $N\approx 10^{23}$ electrons, the relevant amount, $\Delta T_z/N$, is of the order of $10^{-21} \unit{eV}$, suggesting that this experiment could put very stringent bounds on the free parameters of the model. We should keep in mind that when looking at these bounds there is no \textit{a priori} knowledge about the value of the exponents $c^{(\pm,l)}$, thus, the bounds that could be obtained for $\Delta\xi^{(l,m)}$ will have a very large range depending on the level of Planck suppression characterized by these exponents. For a further improvement on the bounds one might want to consider ``moving'' the gravitational sources at the Laboratory in a way that its frame-dragging effect at the location would be larger than the one caused by the Earth's rotation. 

In any event, what is clearly shown here is the viability of experimentally testing for these kind of exotic effects tied to an hypothetical granularity of spacetime in a world governed by perfectly Lorentz invariant laws. Finding a signal would be a fantastic achievement. Not finding any should help in setting restrictions on theories assigning for such granular features to spacetime.

\section*{Acknowledgments}

We want to thank E.G. Adelberger for suggesting the use of the E\"ot-Wash torsion balance for this experiment and for a very useful discussions about the model. This work was presented at the VIII School of the Gravitation and Mathematical Physics Division of the Mexican Physical Society thanks to the PAEP-UNAM 2009 funding project. The work was supported in part by CONACYT 101712 and PAPIIT-UNAM IN119808-3 projects.

\end{document}